\documentclass[12pt,preprint]{aastex}
\usepackage{epsfig}
\usepackage{rotating}
\usepackage{graphicx}

\newcommand{\bc}{\begin{center}}
\newcommand{\ec}{\end{center}}
\newcommand{\be}{\begin{equation}}
\newcommand{\ee}{\end{equation}}
\newcommand{\ba}{\begin{eqnarray}}
\newcommand{\ea}{\end{eqnarray}}
\newcommand{\bt}{\begin{tabular}}
\newcommand{\et}{\end{tabular}}

\newcommand{\chan}{{\sl Chandra}}

\begin{document}

\title{
The Compact Central Object in the
Supernova Remnant G266.2--1.2}

\author{Oleg Kargaltsev, George G.\ Pavlov, Divas Sanwal,
and Gordon P.\ Garmire} \affil{Dept.\ of Astronomy and
Astrophysics, The Pennsylvania State University, 525 Davey Lab,
University Park, PA 16802}
\email{green,pavlov,divas,garmire@astro.psu.edu}

\begin{abstract}

We observed the compact central object CXOU J085201.4--461753 in
the supernova remnant G266.2--1.2 (RX J0852.0--4622) with the
\chan\/ ACIS detector in timing mode. The spectrum of this object
can be described by a blackbody model with the temperature
$kT=404\pm 5$ eV and radius of the emitting region $R=0.28\pm0.01$
km, at a distance of 1~kpc. Power-law and thermal plasma models do
not fit the source spectrum. The spectrum shows a marginally
significant feature at 1.68 keV. Search for periodicity yields two
candidate periods, about 301 ms and 33 ms,
both significant at a 2.1$\sigma$ level; the corresponding
pulsed fractions are 13\% and 9\%, respectively. 
We find no
evidence for long-term variability of the source flux,
nor do we find extended
emission around the central object. We suggest that CXOU
J085201.4--461753 is similar to CXOU J232327.9+584842, the central
source of the supernova remnant Cas A. It could be either a
neutron star with a low or regular magnetic field, slowly
accreting from a fossil disk, or, more likely, an isolated neutron
star with a superstrong magnetic field. In either case, a
conservative upper limit on surface temperature of a 10 km radius
neutron star is about 90 eV, which suggests accelerated cooling
for a reasonable age of a few thousand years.
\end{abstract}

\keywords{stars: neutron --- supernova remnants: individual (G266.2--1.2)
--- X-rays: individual (CXOU J085201.4--461753)}

\section{Introduction}

The shell-like supernova remnant (SNR) G266.2--1.2 (also known as
RX~J0852.0--4622, or ``Vela Junior'') at the south-east corner of
the Vela SNR was discovered by Aschenbach (1998) in the {\sl
ROSAT} All-Sky Survey data. Possible detection of the 1.156 MeV
$\gamma$-ray line of the radioactive isotope $^{44}$Ti (half-life
$\sim 90$ yr) with the Compton Gamma-Ray Observatory (Iyudin et
al. 1998) may imply a very young SNR age of $\sim 680$ yr, at a
distance of $\sim 200$ pc. Aschenbach, Iyudin, \& Sch\"onfelder
(1999) estimated upper limits of 1100 yr for the age, and 500 pc
for the distance. Observations with {\sl ASCA} (Tsunemi et al.
2000; Slane et al. 2001) demonstrate that the X-ray spectra of the
SNR shell are nonthermal. Fits of these spectra with a power-law
(PL) model yield a hydrogen column density substantially higher
than that for the Vela SNR, implying a plausible distance to the
remnant of 1--2 kpc, and an age of a few thousand years.

Aschenbach (1998) suggests that G266.2--1.2 was created by a core-collapse
supernova that left a compact remnant ---
a neutron star (NS) or a black hole (BH). Three compact remnant candidates
have been reported from the
observations with {\sl ROSAT} (Aschenbach 1998; Aschenbach et al.\ 1999),
{\sl ASCA} (Slane et al.\ 2001), and {\sl Beppo-SAX} (Mereghetti 2001).
Pavlov et al.\
(2001) observed G266.2--1.2 with the \chan\ Advanced CCD Imaging
Spectrometer (ACIS)
and found only one bright X-ray source,
CXOU\, J085201.4--461753 (J0852 hereafter),
close to the SNR center.
They measured the source position
with accuracy better
than $2''$ and proved that J0852 is not an X-ray counterpart of
bright optical stars in the field.
Follow-up optical observations
(Pavlov et al.\ 2001; Mereghetti, Pelizzoni, \& De Luca 2002a)
revealed an
object located only $2\farcs4$ south-west of the J0852.
The colors
of the optical source  are consistent with those of a main sequence
star at a distance of 1.5--2.5 kpc; most likely, this
is a  field star.
The limiting optical magnitudes at the position of the
X-ray source ($B>22.5$, $R>21$ --- Pavlov et al.\ 2001;
$B>23$, $R>22.5$ --- Mereghetti et al.\ 2002a)
rule out the possibility that the X-ray source is an AGN.
The lack of variability combined
with the X-ray spectral properties makes a cataclysmic variable
interpretation also implausible. The nature of the source
remains elusive, although an isolated cooling NS or a
NS with a ``fallback'' disk seem to be
possible interpretations.

The large frame time, 3.24 s,
of the previous snapshot (3 ks) ACIS observation
made it impossible to search for short periods and led to strong saturation
(pile-up) of the source image, precluding an accurate spectral analysis.
To search for  pulsations from the compact source and obtain a more accurate
spectrum, we observed J0852 with  {\sl Chandra} ACIS
with a time resolution of 2.85 ms.
We present the
results of
this observation in \S2
and discuss the nature of the source in \S3.

\section{Observation and data analysis}

J0852 was observed with ACIS-S3
in Continuous Clocking (CC) mode on 2001 August 30 (31.5 ks total
exposure).
CC mode allows one to achieve time resolution of 2.85 ms
at the expense of spatial information in one dimension.
There were no substantial
``background flares'' during the observation,
so we do not exclude any time intervals from the analysis.
For data reduction and analysis, we used CIAO 2.2.1 (CALDB 2.7) and
XSPEC v.11.0.

The FWHM of the one-dimensional (1-D) source image is $0\farcs7$,
consistent with the ACIS point spread function. No evidence for 
excess emission around the point source is seen above the background
of 0.013 counts s$^{-1}$ per $1''$ segment of the 1-D image
(equivalent to an average surface brightness of 
0.025 counts ks$^{-1}$ arcsec$^{-2}$).

\subsection{Spectral Analysis}

For the spectral analysis,
we extracted 11,450 source-plus-background counts
from a $4''$ segment of the 1-D image.
The background was
taken from two adjacent $10''$ segments.
The background-subtracted source count rate is $0.313 \pm 0.004$ counts
s$^{-1}$. Figure~1 shows the pulse-height spectrum in the 0.6--8.0 keV
band,
grouped into
77 bins
with $\geq 100$ source
counts per bin.  We ignored all counts below 0.6 keV for spectral
fitting because of the poorly known  ACIS response at lower energies.

Fitting the spectrum with a power-law (PL) model yields a large
photon index $\gamma=4.32 \pm 0.06$ (all uncertainties at a
$1\sigma$ confidence level), and a hydrogen column density
$n_{\rm{H,21}}\equiv n_{\rm{H}}/10^{21}$ cm$^{-2}=11.2\pm 0.2$,
close to the total Galactic HI column density in this direction,
$\approx 1 \times 10^{22}$ cm$^{-2}$ (Dickey \& Lockman 1990;
estimated with the W3NH
tool\footnote{\tt{http://heasarc.gsfc.nasa.gov}}). The quality of
the fit is so poor ($\chi^{2}_{\nu}=3.94$ for 74 degrees of
freedom [d.o.f.]) that this model can be rejected. Thermal plasma
emission models (thermal bremsstrahlung and {\tt mekal} with solar
abundances) also do not fit the observed spectrum
($\chi^{2}_{\nu}=1.63$ and  14.26 for 74 d.o.f., respectively).

On the contrary, a single blackbody (BB) model fits the spectrum reasonably
well ($\chi^{2}_{\nu}=1.13$ for 74 d.o.f.; see Fig.\ 1).
It yields a temperature
$T = 4.68\pm 0.06$ MK ($kT = 404\pm 5$ eV) and a radius
of equivalent emitting sphere
$R=(0.28\pm0.01)\, d_{1}$ km, where
$d_{1}\equiv d/1~\rm{kpc}$. The bolometric luminosity is
$L_{\rm{bol}}=(2.5 \pm 0.2)\times 10^{32}d_{1}^{2}$ ergs s$^{-1}$.
The hydrogen column density, $n_{\rm H,21}=3.45\pm 0.15$, considerably exceeds
the highest value, $n_{\rm H,21}=0.6$, found by Lu \& Aschenbach (2000)
for the Vela SNR. It indicates that the source is substantially more distant
than the Vela pulsar ($d_{\rm Vela}=294^{+76}_{-50}$ pc --- Caraveo et al.\ 2001).
Adding a PL component to the BB model only marginally improves
the fit ($\chi_\nu^{2}=1.126$ for 72 d.o.f., vs.\ 1.130 for 74 d.o.f,
for a single BB ). The
F-test shows that the reduction of $\chi^{2}$ caused by adding the PL
component is significant only at a 66$\%$ confidence level.

Fits with the magnetic hydrogen NS atmosphere models (Pavlov et al.\
1995) give a lower effective
temperature ($kT\approx 270$ eV) and a larger emitting area
($R\approx 1.2 d_1$ km).
In both BB and H atmosphere fits, the inferred radius is much smaller
than the expected NS radius, and the temperature is too high
to interpret the detected X-rays as emitted from the
whole surface of a  uniformly heated isolated NS of a reasonable age.

To constrain the temperature of the entire NS surface, we fit the
spectrum with a two-component BB model. The fits to
the {\sl ASCA} spectra of the outer, brighter parts of the SNR give a range of hydrogen
column densities from $1.4$ to $5.3 \times 10^{21}$ cm$^{-2}$
(Slane et al.\ 2001). To find a conservative upper limit on the
surface temperature $T_{\rm s}$, we fix the column density at $n_{\rm{H},21}=5.3$,
add a soft BB component with $R_{\rm s}=10\,d_1$ km,
and fit $T$ and $R$ at different values of $T_{\rm s}$, increasing $T_{\rm s}$
until the fit probability falls to 0.1\%.
This gives an
upper limit $T_{\rm s} \le 89$ eV,
at a 99.9\%
confidence level. If we fix the column density at $n_{\rm{H},21}=3.4$
(as obtained for the single BB fit), the limit becomes as low as
$T_{\rm s}\le 75$ eV.

Although we find no strong spectral lines, there is a hint of a
spectral feature at 1.68 keV (see Fig.\ 1). This feature
persists when the data are rebinned with different numbers of counts
per bin. We see no anomalies in the data which could explain
the feature as an artifact. In particular, we
have ruled out that the feature
could be caused by anomalously high values in the bias map for S3 chip
(P.\ Ford 2002, private communication).
The shape of the feature resembles the so-called inverse P-Cygni
profile, which might be associated with accretion. Its width, 
$\sim 100$ eV, might correspond to velocities
of accreting material $\sim 0.03 c$;
however, this width is comparable with the spectral
resolution of ACIS-S3 around 1.7 keV (\chan\ Proposers'
Observatory Guide, v.3.0, \S 6.7). 
The feature 
is not seen in the background
(SNR) spectrum.

\subsection{Timing analysis}

For timing analysis, we extracted 10,957 photons from a $2\farcs5$ segment
centered on J0852
($\geq$ $89\%$ of these counts
are expected to come from the point source). The time span
of the observation is $T_{\rm span}=31.5$~ks.
We corrected the event
times
for telescope dither and Science Instrument Module motion
using the approach described by Zavlin et al.\ (2000).
We transformed the corrected times to
the solar system barycenter
using the {\tt axBary} tool of CIAO.

We  used  the $Z_{m}^{2}$ test (Buccheri et al. 1983) to search
for periodic pulsations. We calculated $Z_{m}^{2}$ for $m=1$--5
(where $m$ is the number of harmonics included) at $10^{8}$ equally
spaced frequencies $f$ in the 0.001--100 Hz range. This
corresponds to oversampling by a factor of about 30, compared to
the expected width of $T_{\rm span}^{-1}\approx 30$ $\mu$Hz of the
$Z_{m}^{2}(f)$ peaks, and guarantees that we miss no peaks. The
two most significant peaks we found are  at $f=3.324231\, {\rm
Hz}\pm 3\, \mu{\rm Hz}$ ($P\approx 301$ ms) and $f= 30.369484\,
{\rm Hz}\pm 2\, \mu{\rm Hz}$ ($P\approx 33$ ms)\footnote{The
frequency uncertainties, at a 90\% confidence level,  are
estimated using the method of Gregory \& Loredo (1996); see also
Zavlin et al.\ (2000).}. The most significant $Z_{m,{\rm max}}$
values, $Z_{4,{\rm max}} = 52.9$ for the 301 ms period and
$Z_{1,{\rm max}} = 36.7$ for the 33 ms period, correspond to
96.7\% and 96.8\% significance levels, respectively, for the
number of independent trials $\mathcal{N}=f_{\rm max}
T_{\rm{span}}\approx 3\times 10^{6}$.

The pulsed fractions obtained from the pulse profiles are
$13\%\pm3\%$ and $9.1\%\pm2.5\%$ for the
301 ms and 33 ms
period candidates, respectively. Because of the low significance,
we consider 13\% as an upper limit for the pulsed fraction.

To search for variability on larger time scales, we binned the
data into 200~s bins.
Using the Kolmogorov-Smirnov
test, the hypothesis that the observed numbers of counts in the
bins come from a Poisson distribution (with the mean of 69.756
counts per bin) can not be rejected at a $70\%$ confidence level. We have
also used
the Fourier transform
and found no
periodic signal with a pulsed fraction larger than 12\% in 1--10
mHz frequency range. Therefore, we find no evidence for long-term
variability in the data.

\section{Discussion}

The X-ray data and optical limits indicate that J0852 is the compact remnant
(NS or BH)
of the supernova explosion. The X-ray spectral properties
and the lack of 
radio
emission (Duncan \& Green 2000)
suggest that J0852 is not an active pulsar.
Furthermore, the {\sl Chandra} observations show no 
sign of a pulsar-wind nebula (PWN) around the point source.
From the 3 ks observation in Timed Exposure mode (Pavlov et al.\ 2001),
the $3\sigma$ upper limit on the PWN brightness (in counts arcsec$^{-2}$)
can be estimated as $3(b/A)^{1/2}$, where $b=0.029$ counts arcsec$^{-2}$
is the background surface brightness, and $A$ is the (unknown) PWN area.
Scaling the area as $A=1000 A_3$ arcsec$^2$ (which corresponds to the
transverse size of about $5\times 10^{17} A_3^{1/2}$ cm)
and assuming a PL spectrum with a photon index $\gamma=1.5$--2, we obtain
an upper limit of 
(1.3--$2.0)\times 10^{30} A_3^{1/2} d_1^2$ erg s$^{-1}$ on the PWN
luminosity in the 0.2--10 kev band, for $n_{\rm H, 21}$ in the range
of 1.4--5.3.

The observational
properties of J0852 strongly resemble those of the other radio-quiet
central compact objects (CCOs) in SNRs (see Pavlov et al.\ 2002a
for a review), particularly the CCO in the SNR Cas A
(Murray et al.\ 2002, and references therein).
At least one of these sources, 1E 1207.4--5209, has been proven to be
a NS rotating with a period of 424 ms (Zavlin et al.\ 2000; Pavlov et
2002b).
A number of
possible interpretations of CCOs
have been recently discussed by several authors (e.g. Pavlov et
al.\ 2000, 2001, 2002a; Chakrabarty et al.\ 2001).
The limits on X-ray-to-optical flux ratio for J0852 and the Cas A CCO
virtually rule out models which involve accretion onto a NS or a BH
from a binary companion. If these are accreting objects,
a more plausible source of accreting matter
might be a ``fossil disk'', left over after the SN explosion (van
Paradjis, Taam, \& van den Heuvel 1995). Alternatively, thermal emission from an
isolated, cooling NS could explain the observational results.
We discuss these two options below.

\subsection{Accretion-powered X-ray pulsar?}

If J0852 is an accreting NS,
the observed luminosity, $L_{\rm x}\sim 2\times 10^{32} d_1^2$ erg s$^{-1}$,
could be due to a rather low
accretion rate, $\dot{m} \sim 1.5\times 10^{12} R_6 M_1^{-1} d_1^2$ g~s$^{-1}$,
where $R_6=R_{\rm NS}/(10^6\, {\rm cm})$, $M_{1}=M/M_\odot$.
The accreting matter could be supplied from a fossil (``fallback'') disk.
The formation of
such a disk from the ejecta
produced by a SN explosion was discussed by a number of authors
(e.g. Marsden, Lingenfelter, \& Rothschild 2001, and references
therein). Some models
suggest that a fossil disk can be formed
several days after the SN explosion (``prompt'' disk) and range
from $0.001M_{\odot}$ to $0.1M_{\odot}$, while others
suggest that the
disk can be formed later, years after the SN
explosion (``delayed'' disk). The details of the formation mechanism and the disk
properties are highly uncertain, and, consequently, the accretion
rate $\dot{m}$ is also poorly constrained, but the required value of $\sim 10^{12}$
g s$^{-1}$ is low enough not to exhaust the disk at any reasonable age of J0852.

The accretion onto a NS can proceed in two different regimes
(e.g., Frank, King, \& Raine 1992), depending on the relation
between the corotation radius, $R_c=1.5\times 10^8 P^{2/3}
M_1^{1/3}$ cm, and the magnetospheric radius,
$R_M=3.5\times10^{9}B_{12}^{4/7}\dot{m}_{12}^{-2/7}
M_1^{-1/7}R_6^{12/7}$ cm, where $P$ is the NS spin period,
$B=10^{12} B_{12}$ G is the magnetic field at the NS surface, and
$\dot{m}_{12}=\dot{m}/(10^{12}\, {\rm g}\, {\rm s}^{-1})$. If
$R_{M}>R_{c}$, the infalling material is stopped at the
magnetospheric radius and  expelled as a wind due to 
centrifugal force. In this ``propeller regime'' (Illarionov \&
Sunyaev 1975), X-ray emission is mainly due to optically thin
thermal bremsstrahlung produced in the flow (Wang \& Robertson
1985). Since the thermal bremsstrahlung model does not fit the
observed spectrum, we consider this case unlikely.

If $R_{M}<R_{c}$ (i.e.,
$P\gtrsim 10^2 B_{12}^{6/7} \dot{m}_{12}^{-3/7} M_1^{-5/7} R_6^{18/7}$ s,
or $B\lesssim 4\times 10^9 P^{7/6} \dot{m}_{12}^{1/2} M_1^{5/6}R_6^{-3}$ G),
the accreting matter is able to reach the NS surface.
At extremely low magnetic fields,
$B \lesssim 6\times 10^5 \dot{m}^{1/2} M_1^{1/4} R_6^{-5/4}$ G,
when the magnetospheric radius is smaller than the NS radius,
a hot layer is formed at the boundary between the accretion disk
and the NS surface (e.g., Frank et al.\ 1992).
Since this boundary layer is expected to be optically thin at $\dot{m}\ll 10^{16}$
g~s$^{-1}$ (Inogamov \& Sunyaev 1999), its radiation cannot explain the observed
BB spectrum.
At reasonable magnetic fields,
$B\gg 10^6 \dot{m}^{1/2} M_1^{1/4} R_6^{-5/4}$ G
($R_M\gg R_{\rm NS}$), the accretion flow
is channeled onto the NS poles, producing hot spots of radius
$a\sim R_{\rm NS}^{3/2}/R_{M}^{1/2} \sim 0.17 B_{12}^{-2/7}
\dot{m}_{12}^{1/7} M_1^{1/14} R_6^{9/14}$ km. The observed size
and temperature of the BB-like radiation
are consistent with being emitted from such a cap at $B\sim 10^{11}$ G.
Such an estimate requires a
pulsar period $P\gtrsim 10$ s, much longer than our candidate
periods. If we assume $P=301$ ms, the condition $R_M<R_c$ requires
$B\lesssim 10^9\, \dot{m}_{12}^{1/2} M_1^{5/6} R_6^{-3}$ G and
$a\gtrsim 1.2\, \dot{m}_{12}^{1/7} M_1^{1/14} R_6^{9/14}$ km,
considerably larger than the size of emitting region, $R\approx 0.3 d_1$ km,
inferred from the BB fit.
However, given the crudeness of the polar cap size estimate, which
can be much smaller than adopted above (see, e.g., Frank et al.\ 1992,
and references therein),
we cannot rule out the
candidate period of 301 ms based on the apparent inconsistency between $a$ and $R$.
Thus, in the accretion
hypothesis, J0852 could be a low-luminosity
X-ray pulsar, presumably with a magnetic field
much lower than those of binary X-ray pulsars, slowly accreting
from a fossil disk.
An argument against this interpretation is a lack of nonperiodic
variability in the radiation from J0852, which is commonly
observed from accreting sources (at least, X-ray binaries).
On the other hand, variability could be found in further observations
of this source.
A direct confirmation of the accreting hypothesis would be detection
of an accretion disk, which would require deep IR-optical
observations with high angular resolution.

\subsection{Isolated cooling neutron star?}
One can also assume that J0852 is an isolated (non-accreting) NS
emitting thermal radiation from its surface.
The ``standard'' NS cooling models
predict a luminosity of $\sim (0.5$--$2)\times 10^{34}$
erg s$^{-1}$
for a NS of 0.1--10 kyr age
(e.g., Tsuruta 1998).
The lower
observed luminosity of J0852 could be interpreted as
due to an accelerated cooling mechanism, but applicability
of the cooling models to J0852 is questionable because
the models assume a uniformly heated NS surface while
the size of the emitting region obtained from
the BB fit is only $\approx 0.3\, d_{1}$ km.

Apparent sizes of the emitting regions much smaller than the
canonical NS radius have been observed from other isolated NSs
(Pavlov et al.\ 2002a,c). In particular, the Cas A CCO shows a
(blackbody)
 size of 0.3 km with a temperature of
 0.6 keV (Pavlov et al.\ 2000), which hints that it is an object
similar to J0852, with a higher temperature possibly due to its younger age.
Pavlov et al.\ (2000) suggested a two-component thermal model for the
Cas~A CCO, in which the observed X-rays are emitted from hydrogen
polar caps of about 1 km radius and 0.24 keV effective temperature,
while the rest of the NS surface is iron at a temperature of 0.15 keV,
too cold to be observable because of strong interstellar absorption.
In this model, the polar caps are hotter
because of
the higher thermal conductivity of hydrogen. Weaker ISM absorption for J0852
allowed us to find a lower temperature limit for the cold component,
$<90$ eV, too low to explain the temperature difference
by different chemical compositions.
It should be mentioned that this limit
is a factor of 1.4 lower than the temperature
predicted by the so-called standard (slow) cooling
model for a $10^{3}$ yrs old NS (see, e.g., Fig.\ 4
in Slane, Helfand, \& Murray 2002). This may indicate that
if J0852 is a NS, it undergoes fast cooling, perhaps associated
with direct Urca processes in the NS core
(e.g., Yakovlev et al.\ 2002).

Hot spots on the NS surface could also be associated with a very strong
magnetic field, $B\gg 10^{13}$ G.
Due to anisotropic heat
conductivity of the NS crust, the surface
temperature is higher at the magnetic poles
(Greenstein \& Hartke 1983; Shibanov \& Yakovlev 1996).
To produce small hot spots, the surface magnetic
field should be strongly nonuniform
(e.g., an offset dipole or a quadrupole ---
Page \&
Sarmiento 1996).
Fast
decay of a superstrong magnetic field ($B\gtrsim 10^{14}$ G)
could provide an additional source of polar cap heating (Thompson \&
Duncan 1996; Colpi, Geppert \& Page 2000).
In such strong magnetic fields,  electron-positron pair creation
should be suppressed due to photon splitting (Baring \& Harding 2001),
which is consistent with the apparent lack of pulsar activity in J0852.

One can crudely
estimate the magnetic field assuming that
one of the two candidate periods, 33 ms or 301 ms, is the true period.
If the initial period of the pulsar was much shorter than the
current period, then the period derivative,
rotation energy loss rate,
and ``canonical'' magnetic field [$B\equiv 3.2\times 10^{19}(P\dot{P})^{1/2}$ G],
can be estimated as $\dot{P}=3.2\times 10^{-11} P[(n-1) t_3]^{-1}$,
$\dot{E}=1.25\times 10^{36} P^{-2} [(n-1)t_3]^{-1}$ erg~s$^{-1}$,
and $B=1.8\times 10^{14} P [(n-1) t_3]^{-1/2}$ G, where
$t=10^3 t_3$ yr is the NS age, and $n$ is the braking index. Assuming $n=2.5$,
(close to that observed in young pulsars), we obtain, for $P=33$ ms,
$\dot{P}=7.0\times 10^{-13} t_3^{-1}$,
$\dot{E}=8.6\times 10^{38} t_3^{-1}$ erg~s$^{-1}$,
and  $B= 4.9\times 10^{12} t_3^{-1/2}$ G --- parameters typical for
a young, active pulsar, in apparent contradiction with observations.
On the other hand, for the 301 ms period, we obtain
$\dot{P}=6.4\times 10^{-12} t_3^{-1}$, $\dot{E}=9.2\times 10^{36} t_3^{-1}$
erg~s$^{-1}$, and $B=
4.4\times 10^{13} t_3^{-1/2}$ G.
Since the local magnetic field can be much higher than the canonical
value (e.g., for an offset dipole), one can speculate that,
for $P=301$ ms,
it is high enough to explain
the hot spot(s) and the
lack of radio-pulsar activity.
If this hypothesis is correct, the J0852 could be
a very young Anomalous X-ray Pulsar (AXP) whose period will become
of order 6--12 s (as observed in AXPs) when
it grows older by a factor of 20--40.

However, there are considerable differences between the properties of AXPs and
J0852.
Contrary to AXPs, whose spectra contain both the BB and PL components
of comparable luminosities (Mereghetti et al.\ 2002b),
the spectrum of J0852 fits well with a single BB model.
The size of the emitting region in J0852
is substantially smaller (0.3 km vs.\ 0.7--5 km), and the temperature somewhat lower
(0.4 keV vs.\ 0.4--0.6 keV), than
those of AXPs.
These differences (particularly, the lack of a PL component
in J0852) hint at different NS parameters.
For instance, it is quite possible that none of the candidate
periods is correct, and the true period is even longer than
the AXP periods.
In this case, the
magnetic field could be even higher than those adopted in the
magnetar interpretation of AXPs --- e.g., $B=2.4\times 10^{15}
(P/20\,{\rm s}) t_3^{-1}$ G.
Such a strong field can inhibit not only the pair cascade, but
also the emission of primary particles from the NS surface,
which might explain the lack of particles in the NS magnetosphere
(hence, the lack of nonthermal radiation) in J0852.
If the NS rotates sufficiently slow, $P
\gtrsim 0.5 B_{15}^{4/15}(Z/26)^{-12/15}$ s,
the critical parallel electric field required to pull out electrons
from the NS surface, $E_{\parallel,\rm{crit}}\approx 2.7 \times
10^{12} (Z/26)^{6/5} B_{15}^{3/5}$ V cm$^{-1}$ (Usov \& Melrose
1995), is higher than the maximum parallel electric field at the surface,
$E_{\parallel,\rm{max}}\approx 1 \times 10^{10}
B_{15}(P/20\,{\rm s})^{-3/2}$ V cm$^{-1}$.
On the other hand, the surface temperature, $kT_e\approx 0.5
(Z/26)^{4/5}B_{15}^{2/5}$ keV, above which the thermoionic
emission of electrons becomes efficient (Usov \& Melrose 1995),
grows with increasing magnetic field.
(These estimates assume
that the NS has no light-element [e.g., hydrogen] atmosphere.)
Therefore, a long period and a superstrong magnetic field
might explain the lack of the PL tail in the
spectrum of J0852 and
other enigmatic CCOs (e.g., in the Cas A and
Pup A SNRs;\  Pavlov et al. 2002a).

In summary, the observations of J0852 can be explained assuming 
it is a NS.
Given the deep limiting
optical magnitudes and the lack of nonperiodic variability, we consider the
interpretation in terms
of an isolated NS with a very strong magnetic field
somewhat more plausible than the accretion models.
Further observations are required to confirm or reject this
hypothesis.
Particularly important would be X-ray timing observations to
measure the period unequivocally,
high-resolution X-ray spectral observations to
look for spectral features, and IR-optical observations to search
for a NS counterpart (e.g., a fossil disk).

\acknowledgements
We thank Glenn Allen and Allyn Tennant for the advice
on the ACIS timing issues, Leisa Townsley and George Chartas
for the advice on the spectral issues,
Peter Ford for examining the bias map, Bing Zhang for useful
discussions, and the anonymous referee for valuable comments.
This work was partly supported by NASA grants NAG5-10865, NAS8-38252, and NAS8-01128.

\clearpage

\begin{figure}[ht]
\includegraphics[width=5.0in, angle=90]{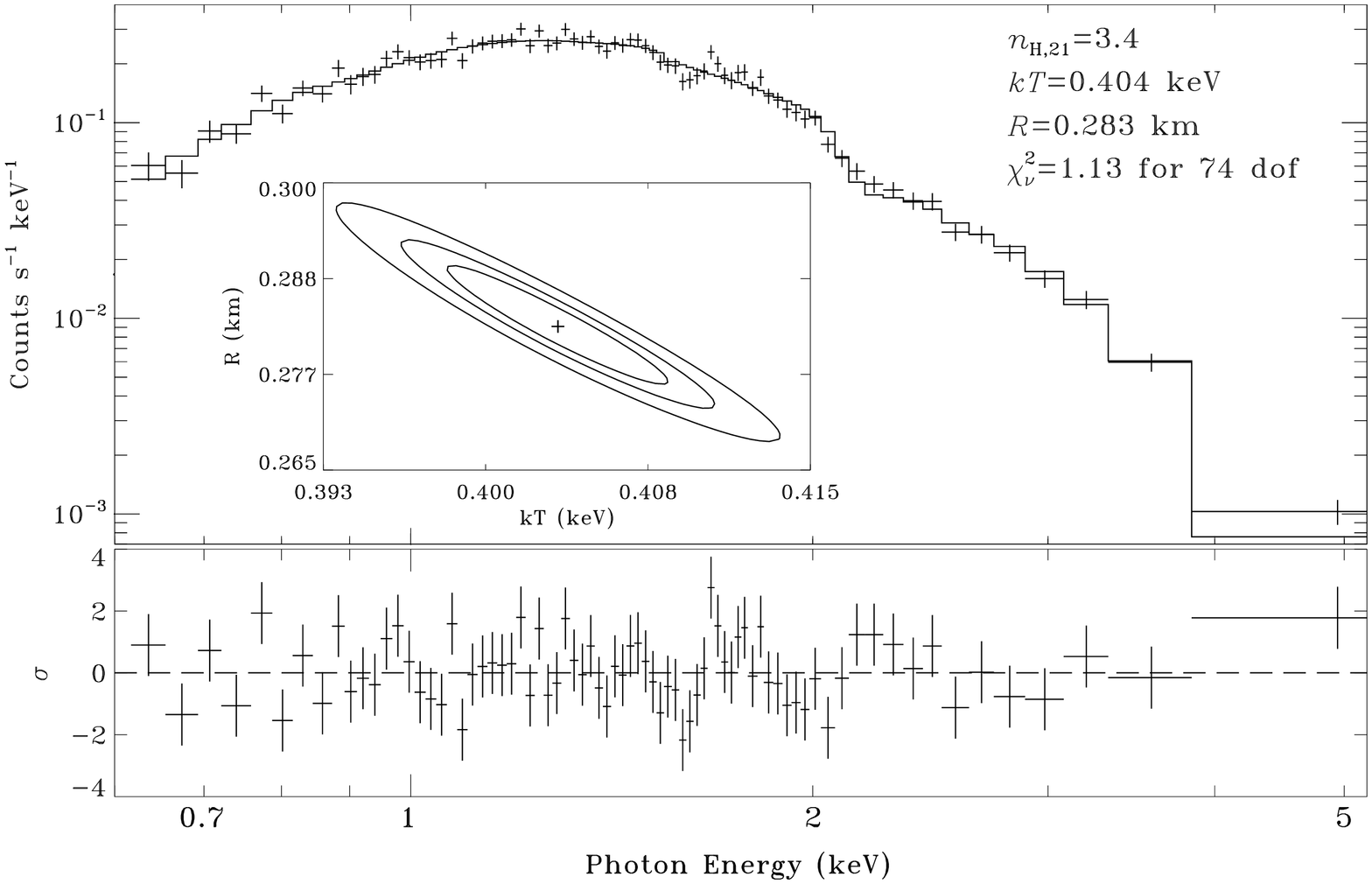}
\caption{ Fit of the ACIS-S3 spectrum of J0852 with a blackbody model.
The contours correspond to 68$\%$, 90$\%$ and 95$\%$ confidence levels.
A possible spectral feature is seen at about 1.68 keV.
}
\label{fig1}

\end{figure}

\end{document}